# A Mitigation Score for COVID-19

Jonathan Cohen
Applied Physics Laboratory
Johns Hopkins University
11100 Johns Hopkins Road
Laurel, Maryland 20723

## 1. Introduction

This note describes a simple score to indicate the effectiveness of mitigation against infections of COVID-19 as observed by new case counts.  The score includes normalization, making comparisons across jurisdictions possible.  The smoothing employed provides robustness in the face of reporting vagaries while retaining salient features of evolution, enabling a clearer picture for decision makers and the public.

## 2. The idea

While the daily number of new cases always has relevance to resource allocation, its current value alone does not say much about the effectiveness of strategies to decrease contagion.  For example, on 15 June 2020, Oregon reported 443 new cases, while the state of New York reported 620.  Are either of those numbers cause for alarm or relief?  It turns out that the largest number of new cases in a single day ever reported by Oregon was, in fact, 443, while the largest number for New York was about 11 thousand, suggesting that Oregon is not doing well at mitigating new infections, while New York appears to be doing a good job, despite the New York number being larger. Clearly, the numbers must be viewed in the context of the past.

This observation suggests a simple measure of mitigation effectiveness: the ratio of the current case rate to the highest observed case rate.  The term "case rate" glosses over the fact that new cases are reported erratically, being subject to batching, errors that are later corrected, and other perturbations.

Below is an overview of the approach and a guide to the sections.

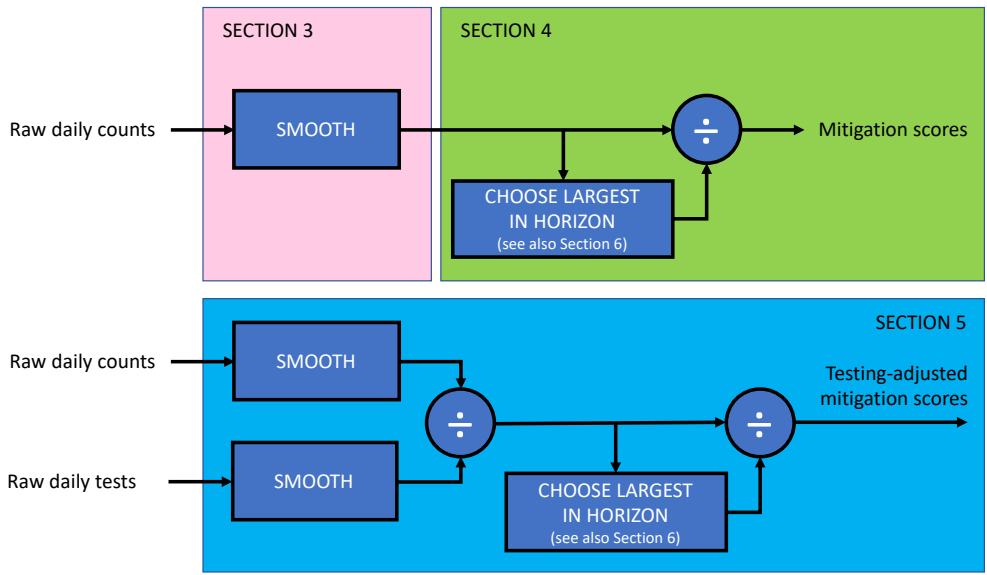

## 3. Smoothing

We are given a set of new case counts $c_1, c_2, \ldots, c_T$, where $c_d$ is the number of new cases reported for day $d$. Note that $T$ is the number of days for which we have observations, and advances every day.

We wish to act on "smoothed" counts $\{\bar{c}_i\}$ that better reflect an underlying case rate and accommodate both delays in reporting and error correction that often occurs within a few days following the errors. That smoothing can be implemented by convolving the raw counts with a window whose weights are $\{w_i\}$:

$$\bar{c}_d = \left(\frac{1}{\sum_{i=d-T}^{d-1} w_i}\right) \sum_{i=d-T}^{d-1} w_i \, c_{d-i},$$

with the parenthesized term normalizing the applied weights to sum to 1. The slightly complicated appearance of expression above is due to the need to accommodate the finite range of times for which we have counts. In particular, the most current smoothed value, $\bar{c}_T$, is given by

$$\bar{c}_T = \left(\frac{1}{\sum_{i=0}^{T-1} w_i}\right) \sum_{i=0}^{T-1} w_i \, c_{T-i}$$

In practice, the window has finite duration, and the summing is done over only a small number of points.

It might be tempting to choose a rectangular window, but that will lead to unnecessary jaggedness and does not reflect the likelihood that relevance of counts to a day in question diminishes more smoothly with distance. A better choice is a window that has a softly rounded center with skirts that fall off more quickly.

Since the details of erratic reporting are not known, it might be hard to claim a particular shape as being "optimum." The author chose a Hamming window, popular in signal processing:

$$w_i = \begin{cases} 0.54 + 0.46 \cos(i\pi/W), & |i| \leq W, \\ 0, & \text{otherwise,} \end{cases}$$

where $W$ is the width. Note that while this computation uses $2W + 1$ days of data, its "effective" width is approximately $W$ days, since $w_i \approx \frac{1}{2}$ when $|i| = \lfloor W/2 \rfloor$.

Figures 1 through 5 show the raw[1] and smoothed daily new case counts for five jurisdictions using a Hamming window of varying width. In particular, Figure 3 shows a correction close to the middle of the period, where the number of new cases is negative, fixing an over-reported count a few days prior. Figure 4 appears to show a positive correction, fixing a day in which no new cases were reported. Figures 1, 2, 3, and 5 show 30 days; Figure 4 shows 90 days.

It would appear that a width of 7 is a good compromise between smoothing uninformative irregularities and preserving salient behavior.

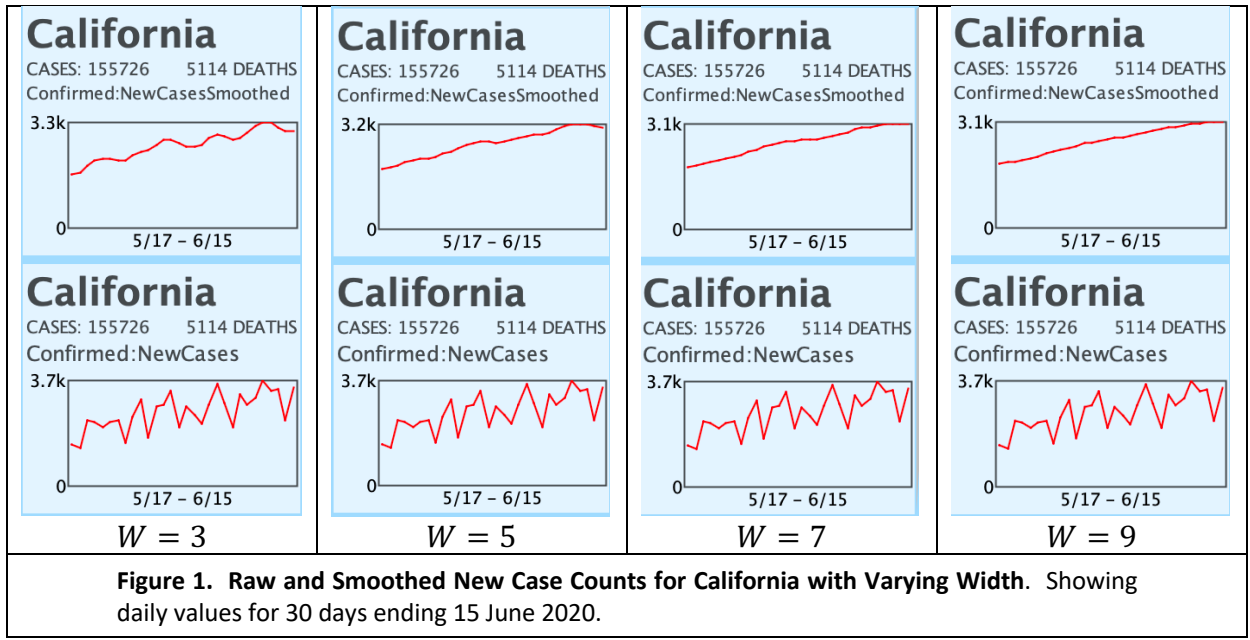

**Figure 1. Raw and Smoothed New Case Counts for California with Varying Width**. Showing daily values for 30 days ending 15 June 2020.

---

[1] All confirmed case data comes from https://raw.githubusercontent.com/CSSEGISandData/COVID-19/master/csse_covid_19_data/csse_covid_19_daily_reports/

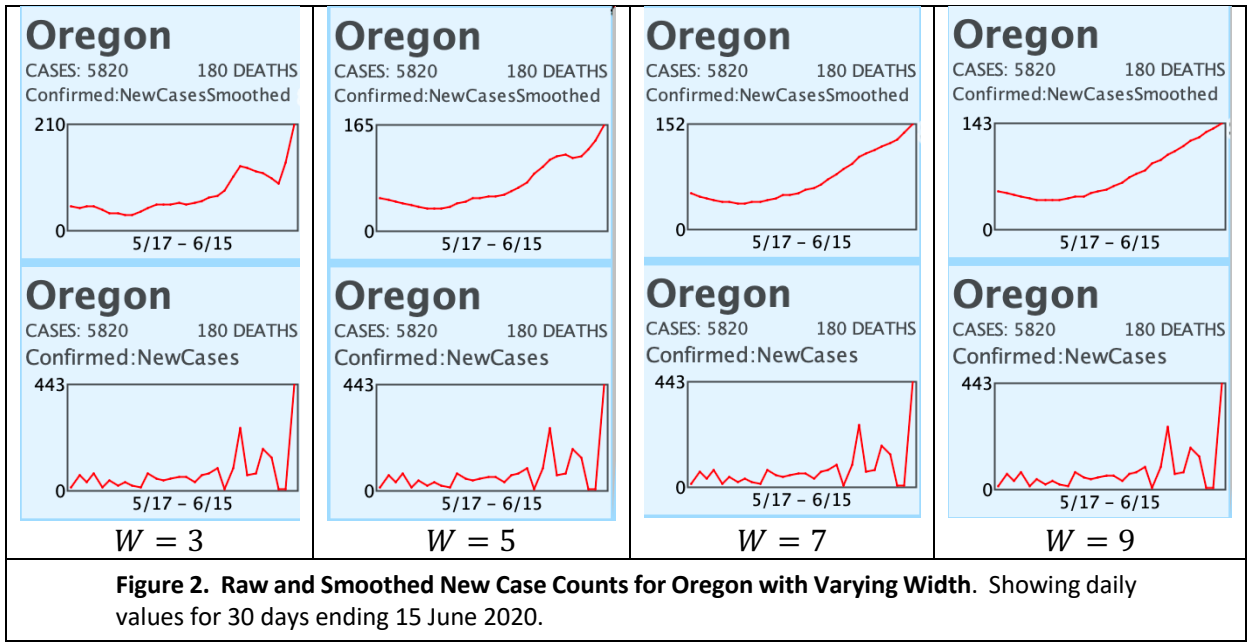

**Figure 2. Raw and Smoothed New Case Counts for Oregon with Varying Width.** Showing daily values for 30 days ending 15 June 2020.

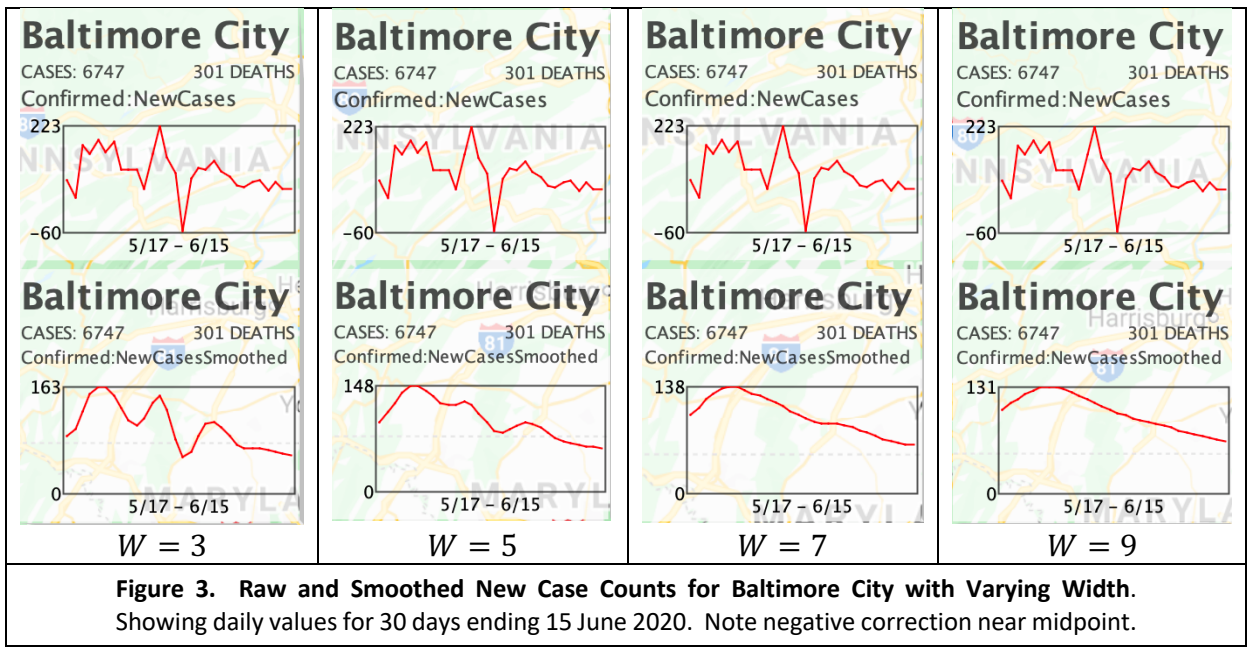

**Figure 3. Raw and Smoothed New Case Counts for Baltimore City with Varying Width.** Showing daily values for 30 days ending 15 June 2020. Note negative correction near midpoint.

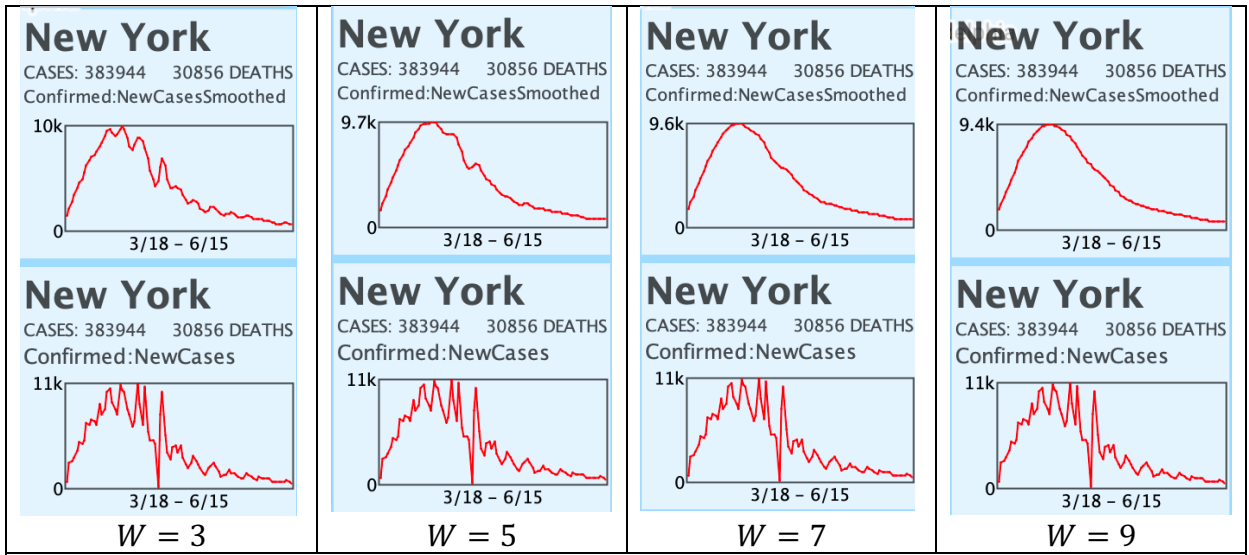

**Figure 4. Raw and Smoothed New Case Counts for New York State with Varying Width.** Showing daily values for **90** days ending 15 June 2020.

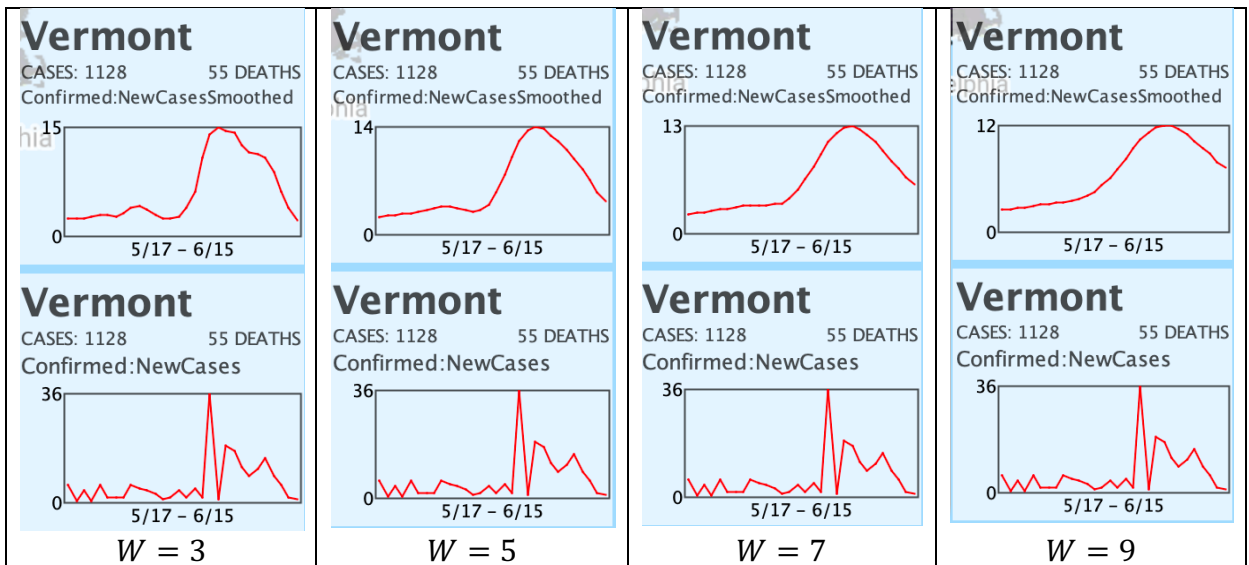

**Figure 5. Raw and Smoothed New Case Counts for Vermont with Varying Width.** Showing daily values for 30 days ending 15 June 2020.

It is useful to compare the suggested Hamming smoothing to the often-employed rectangular windowing, also referred to as a "moving average" or "rolling average" or "running average" or "sliding average". For a rectangular window of width $W$, the value reported on day $d$ is the average value over $W$ consecutive days, either ending on day $d$ (so the window is to the left of the day) or centered on day $d$. In the latter case, the most recent reports can make use only of data up to the present, so they must average over fewer days (just as employed above for the weighted averaging). The most often employed smoothing seems to have settled on a 7-day moving average ending on the day of reporting,[2] though early reporting often used 3-day averages.

Figures 6 and 7 offer a comparison between windowing using the Hamming window ($W = 7$) and simple running averages, employing rectangular windows.  Figure 6 shows running averages centered on the centered on the day being described, while Figure 7 shows a window to the left of the day, that is, the window ends on the day of reporting.  In the latter case, the averaged values lag the raw values by a half of the window size.

The figures show that simple running averages using $W = 3$ are subject to the wild fluctuations common in the reported data.  When the averaging window is increased to having the same effective window size as the Hamming window ($W = 7$), it still exhibits distracting jaggedness.  Even when using $W = 15$, the same number of data points as the Hamming window, the running average suffers uninformative deviations.  (This is more obvious in Figure 7, due to the offset in time.)

---

[2] Examples (including confirmed cases, testing, and deaths): https://covid.cdc.gov/covid-data-tracker/#trends_dailytrendscases, https://coronavirus.maryland.gov, https://www.statista.com/statistics/1111867/trailing-seven-day-average-number-of-covid-19-deaths-select-countries-worldwide/, https://healthdata.gov/dataset/covid-19-daily-rolling-average-case-death-and-hospitalization-rates, https://coronavirus.jhu.edu/data/new-cases

Cohen: A Mitigation Score for COVID-19
16 June 2020; revised 18 November 2020

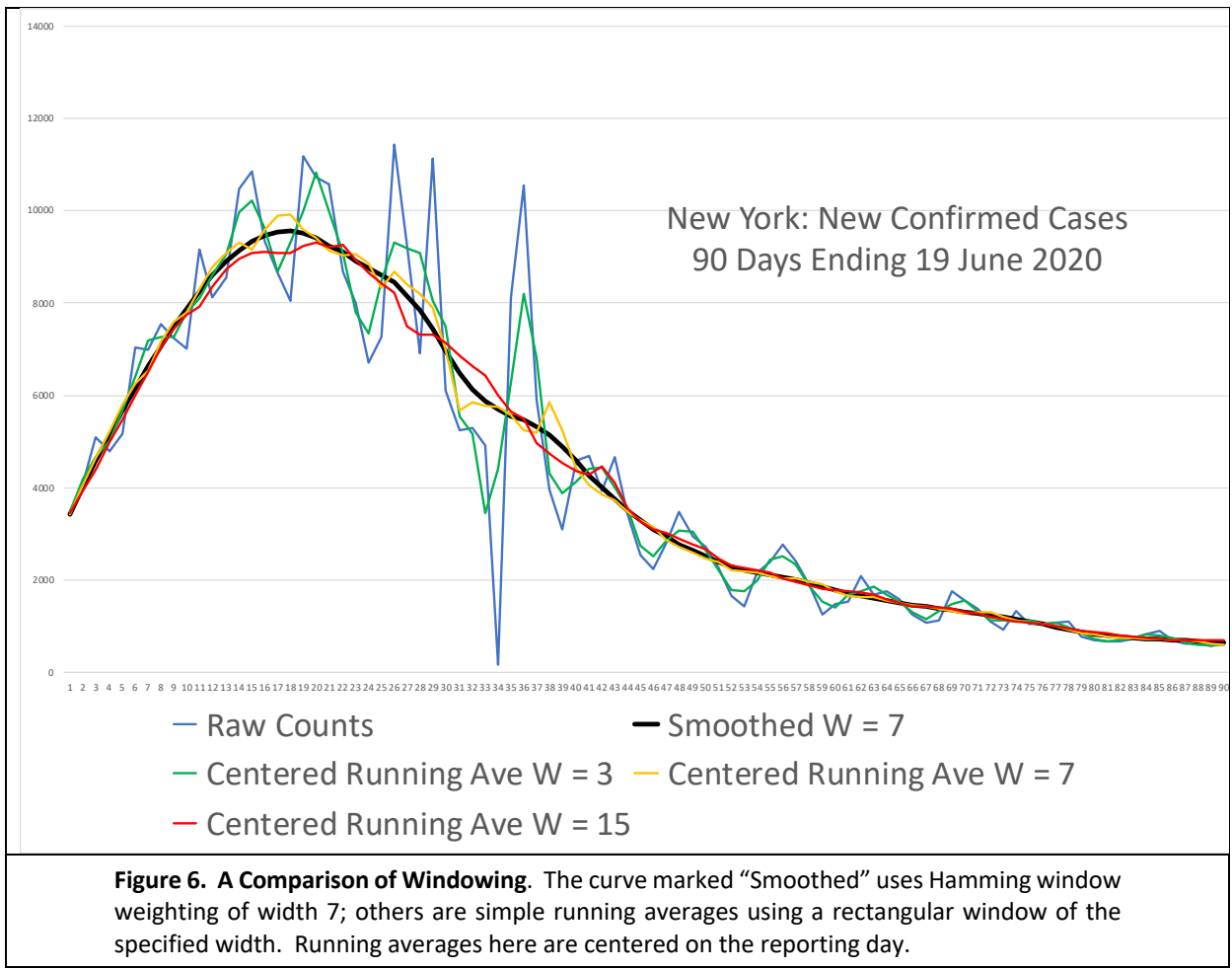

**Figure 6. A Comparison of Windowing.** The curve marked "Smoothed" uses Hamming window weighting of width 7; others are simple running averages using a rectangular window of the specified width. Running averages here are centered on the reporting day.

Cohen: A Mitigation Score for COVID-19　　　　　　　　　　　　　　　　　　　　　　　　　　　8
16 June 2020; revised 18 November 2020

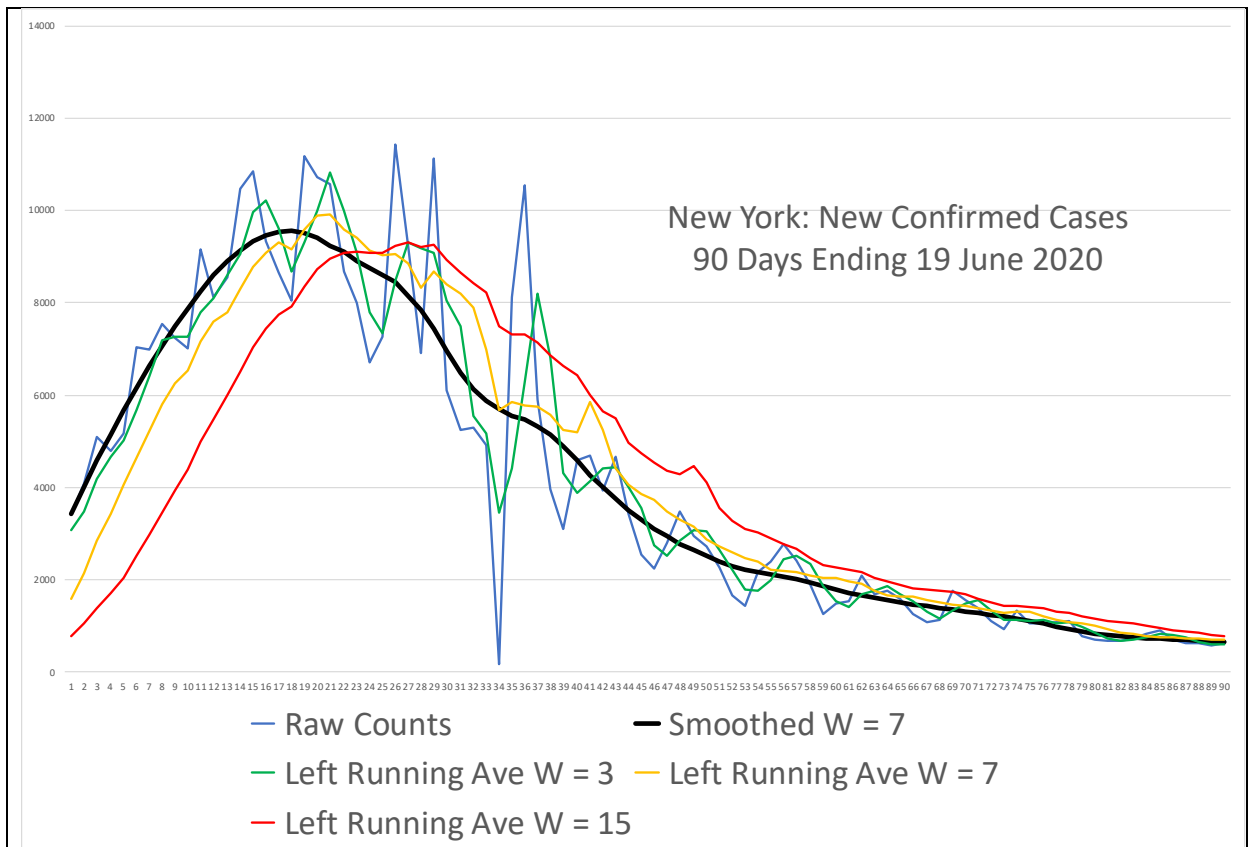

**Figure 7. A Comparison of Windowing.** The curve marked "Smoothed" uses Hamming window weighting of width 7; others are simple running averages using a rectangular window of the specified width. Running averages here use windows that end on the reporting day.

## 4. The Simple Mitigation Score[3]

Having smoothed the counts, we are in a position to produce a mitigation score. The (simple) mitigation score $A_d$ for day $d$ is defined as

$$A(d) = \frac{\bar{c}_d}{\max_{t \in \{1,2,\ldots,d\}} \{\bar{c}_t\}}, \quad d \in \{1, 2, \ldots, T\}.$$

A value near zero means that the current rate is small relative to the maximum rate; a value equal to one means that the current rate is as large as it has ever been.

---

[3] In earlier versions and implementations, this score was referred to as the "abatement" score, a label which also appears in some of the figures here.

Cohen: A Mitigation Score for COVID-19
16 June 2020; revised 18 November 2020

Figure 8 shows a heatmap of the continental United States whose colors indicate the confirmed new cases mitigation score for 15 June 2020, with a value of zero represented by green, a value of one represented by red, and intermediate values shown by intermediate colors. Also shown are the raw new case counts and the mitigation score over the preceding 90 days for the state of New York. Although New York was one of the hardest hit areas, the measures taken there appear to have been quite successful in reducing the spread of infection.

Conversely, Figure 9 shows that while new cases had been reduced in Florida, progress reversed dramatically, resulting in a new high of infection rate.

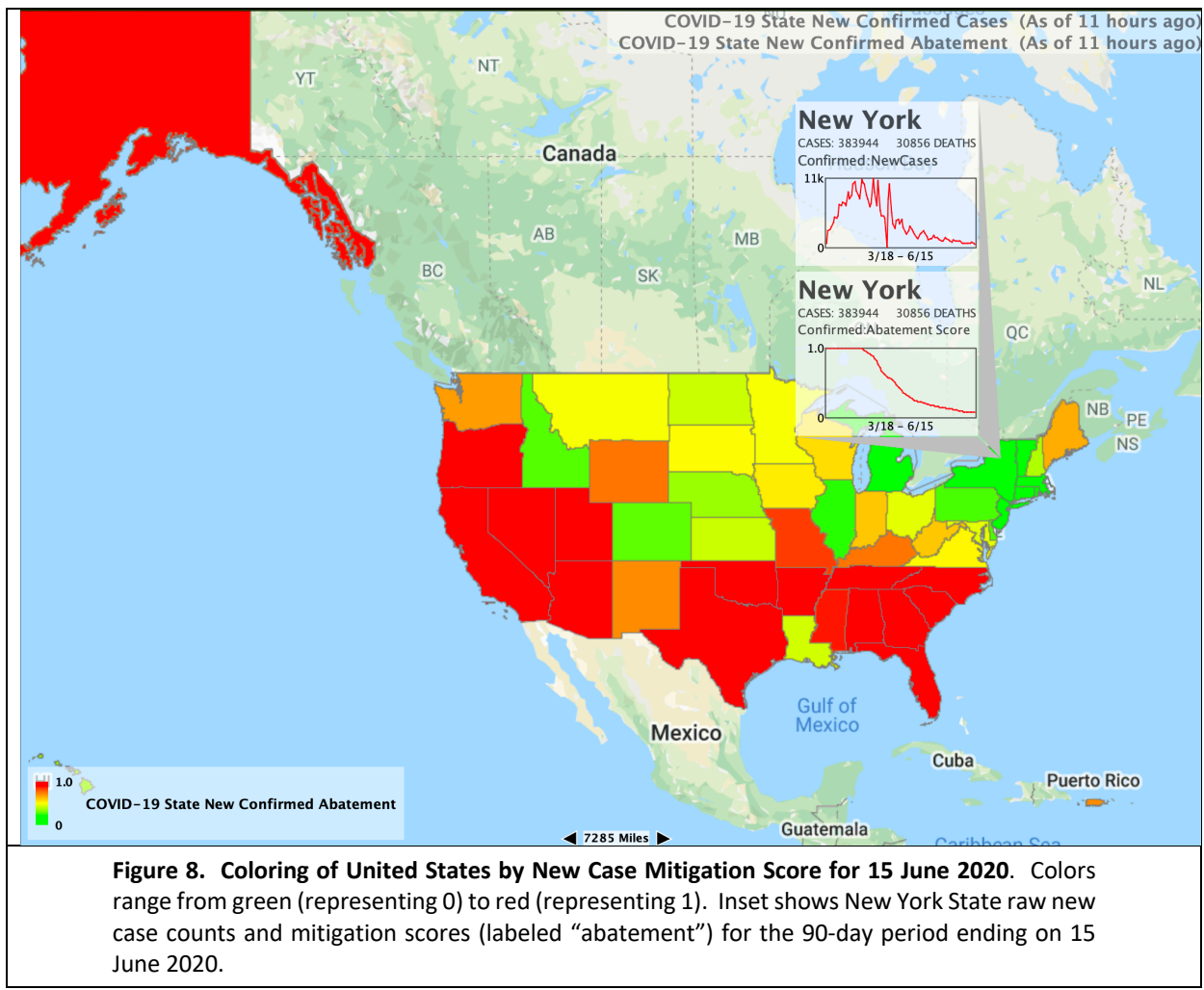

**Figure 8. Coloring of United States by New Case Mitigation Score for 15 June 2020**. Colors range from green (representing 0) to red (representing 1). Inset shows New York State raw new case counts and mitigation scores (labeled "abatement") for the 90-day period ending on 15 June 2020.

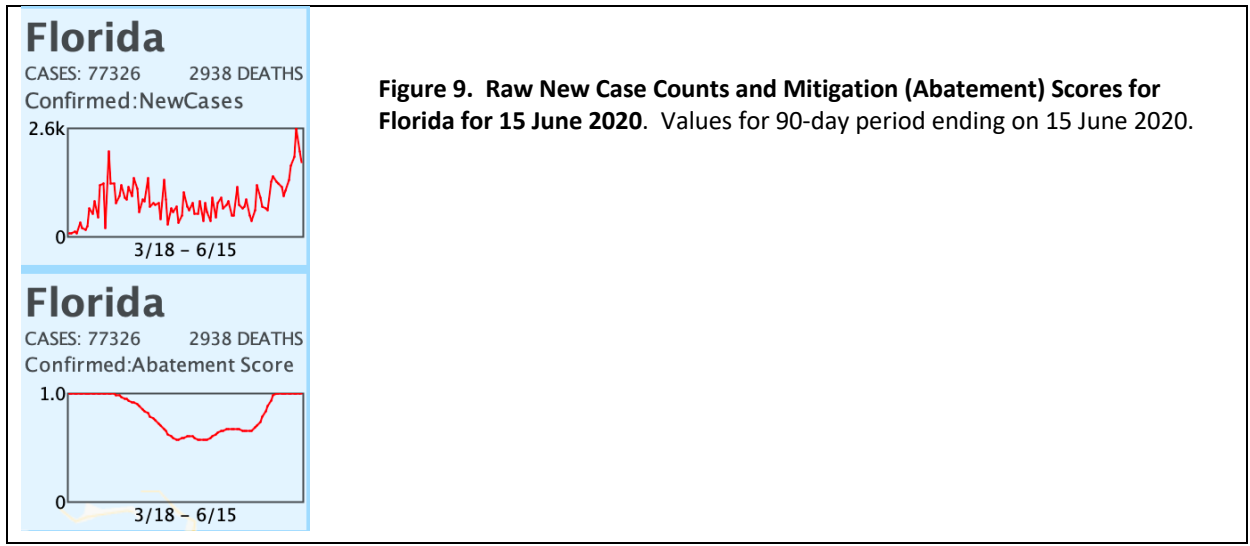

**Figure 9. Raw New Case Counts and Mitigation (Abatement) Scores for Florida for 15 June 2020.** Values for 90-day period ending on 15 June 2020.

## 5. Adjusting for testing

An obvious concern about the formulation of a mitigation score so far is that it would favorably judge a jurisdiction that had stopped reporting altogether. Less drastic would be a lowering of the score simply due to a lessening of testing. Conversely, a jurisdiction that aggressively increased testing would be penalized for the larger number of new cases discovered. Accordingly, if testing numbers are known, they should be incorporated in the score.

Let $p_i$ be the number of tests conducted on day $i$. The reporting of tests $\{p_i\}$ suffers from the same temporal perturbations that the counts $\{c_i\}$ do. We may deal with those irregularities by smoothing the test counts in the same manner as the case counts, using the resulting $\{\bar{p}_i\}$, where

$$\bar{p}_d = \left(\frac{1}{\sum_{i=d-T}^{d-1} w_i}\right) \sum_{i=d-T}^{d-1} w_i \, p_{d-i} \, .$$

While there is no requirement that the same smoothing window be applied to the test counts as the case counts, there seems no motivation to do otherwise, and it will be assumed for examples below that the same Hamming window is applied to both.

To incorporate the testing counts into the mitigation score, we divide the (smoothed) case counts by the (smoothed) testing counts. The testing-adjusted mitigation score for day $d$ is

$$A(d) = \frac{\bar{c}_d/\bar{p}_d}{\max_{t \in \{1,2,\ldots,d\}}\{\bar{c}_t/\bar{p}_t\}} \quad , \quad d \in \{1, 2, \ldots, T\}.$$

Note that setting the testing rate to a constant causes the testing-adjusted mitigation score to reduce to the simple mitigation score defined earlier. Stated another way, if we don't have testing statistics, we use the simple (unadjusted) mitigation score, which carries the implicit assumption that the testing rate is uniform.

Testing-adjusted mitigation is best done with horizon limiting, described below.

## 6. Limiting the horizon

The mitigation score compares the state in the neighborhood of a day in question to the worst state seen in that jurisdiction up to that time. It may be too generous to consider *all* prior times, since so much has changed in the understanding, reporting, testing, and response to the disease. Moreover, we are probably more interested in a more recent behavior. The fact that we are much better than half a year ago may be small comfort if we are much worse than a month ago. Accordingly, it may be appropriate to define a horizon on interest before which we ignore what happened. Let that horizon be specified as being $H$ days long. Then our horizon-constrained mitigation score for day $d$ is normalized by the maximum over the $H$ days ending in day $d$:

$$A_H(d) = \frac{\bar{c}_d/\bar{p}_d}{\max_{t \in \{d-(H-1), d-(H-2),\ldots,d\}}\{\bar{c}_t/\bar{p}_t\}} \quad , \quad d \in \{1, 2, \ldots, T\}.$$

## 7. Examples of testing adjustment

The use of testing-adjusted[4] mitigation with a sensible horizon is revealing, as Figure 10 shows, which compares testing-adjusted 30-day horizon scores to unadjusted scores with no horizon constraint. While many states, mostly in the middle of the country (ND, SD, NE, KS), look less favorable when testing is considered, some states (MN, SC, GA) benefit from testing consideration.

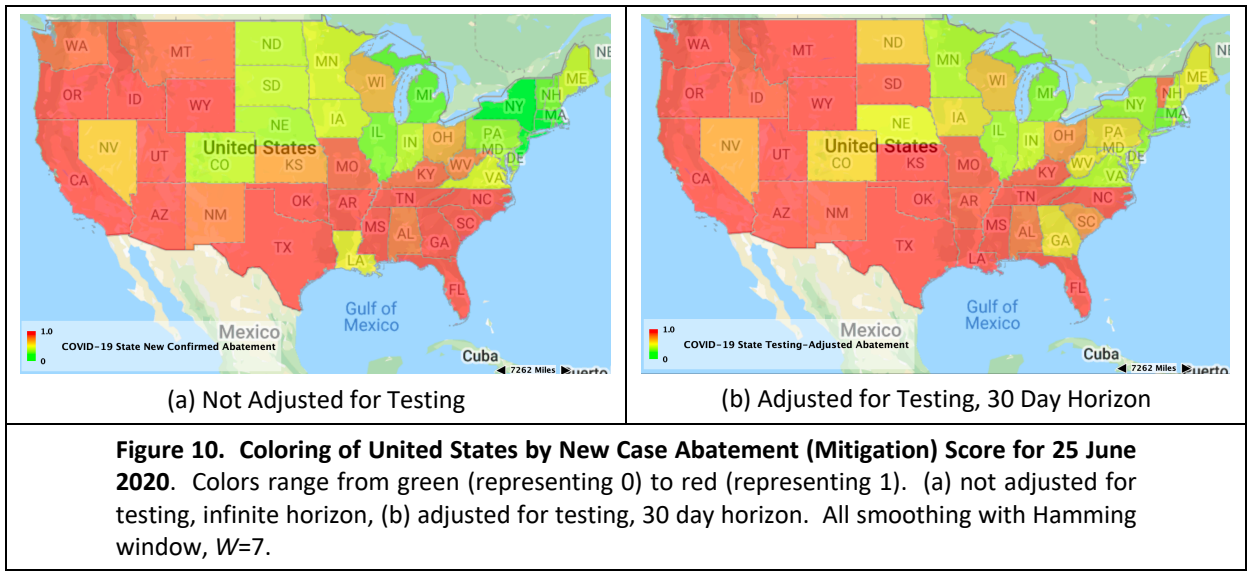

(a) Not Adjusted for Testing  (b) Adjusted for Testing, 30 Day Horizon

**Figure 10. Coloring of United States by New Case Abatement (Mitigation) Score for 25 June 2020.** Colors range from green (representing 0) to red (representing 1). (a) not adjusted for testing, infinite horizon, (b) adjusted for testing, 30 day horizon. All smoothing with Hamming window, $W$=7.

Figure 11 amplifies two of these cases: South Dakota (which looks worse when considering testing rates) and Georgia (which looks better). In particular, South Dakota shows a steady decline in testing over roughly two weeks leading up to 20 June 2020, while Georgia has had growth in testing.

The figure also illustrates the point of limiting the horizon: early reports of testing in many jurisdictions suggest very low numbers, making testing-adjusted statistics misleading. Also, the maxima in the distant past may not matter as much as a more recent history, such as in the case of South Dakota.

---

[4] Testing data was obtained from the COVID Project, https://covidtracking.com/api/v1/ states/daily.json

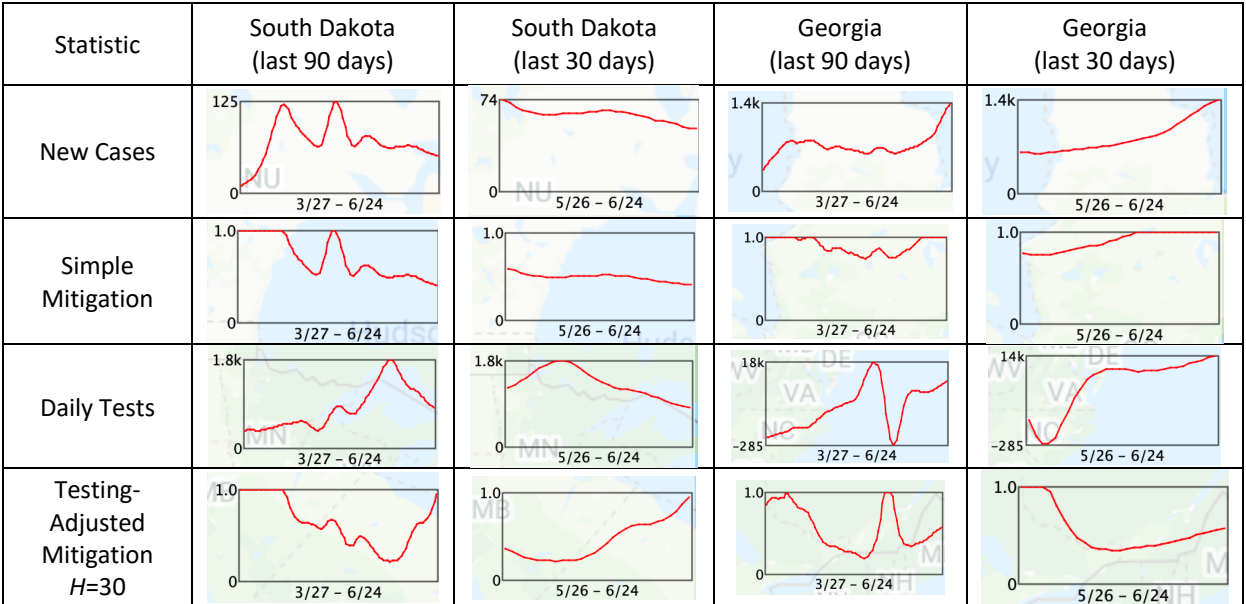

**Figure 11. Comparison of Testing-Adjusted and Simple Mitigation Scores for 25 June 2020.** Charts show the last 90 days or 30 days. All graphs smoothed with Hamming window, *W*=7. The small negative value in Georgia's testing indicates a correction to data, retracting counts from previous days.

## 8. Discussion

The mitigation score suggested here serves as a simple way to convey the effectiveness of mitigation measures. It accommodates comparison of jurisdictions of widely varying size because of local normalization and is tolerant of always-present perturbations in data reporting. Where testing data is available, the simple mitigation score may be augmented, becoming the testing-adjusted mitigation score.

The simple mitigation score may be applied to other statistics such as death counts and testing scores.

One could choose any of a variety of smoothing kernels. The Hamming window was chosen somewhat arbitrarily, but has the requisite slowly varying top with sharper skirts and has smooth, but finite, tails. As Figures 6 and 7 show, this windowing provides some mitigation of distracting reporting irregularities.

The author had initially planned to report the complement of the above definition of the mitigation score, so that a value of 1 meant complete mitigation, and 0 meant no mitigation. But others have pointed out that most people, especially the lay public, have become accustomed to seeing values near zero as good and values near one as bad; hence, the current definition.

## 9. Acknowledgements

The new case data was drawn from the daily snapshots compiled by Johns Hopkins University and found here:

> https://raw.githubusercontent.com/CSSEGISandData/COVID-19/master/csse_covid_19_data/csse_covid_19_daily_reports/

Data on testing was obtained from the COVID Project, and found here:

> https://covidtracking.com/api/v1/ states/daily.json

This note benefited from the suggestions of Alan Ravitz.